\documentclass[iop,showpacs,preprint]{revtex4-1}
\usepackage{amsmath,amssymb,amsfonts}
\usepackage[english]{babel}
\usepackage{graphicx}
\usepackage{dcolumn}
\usepackage{bm}
\usepackage{color}
\usepackage{subfig}

\begin{document}
\title{Chaotic dynamics of thermal atoms in labyrinths created by optical lattices}

\author{R. P\'erez-Pascual$^1$}

\author{B. M. Rodr\'{\i}guez-Lara$^{1,2}$}

\author{R. J\'auregui$^1$}
\affiliation{$^1$Instituto de F\'{\i}sica, Universidad Nacional Aut\'{o}noma de M\'{e}xico,
Apdo. Postal 20-364, M\'{e}xico D.F. 01000, M\'{e}xico}
\affiliation{$^2$Institute of Photonic Technologies, National Tsing-Hua University, Hsinchu 300, Taiwan}

\date{\today}

\begin{abstract}
We study the dynamics of non interacting thermal atoms embedded in structured
optical lattices with non trivial geometry. The lattice would be generated by
two counter propagating modes with parabolic cylindrical symmetry and we concentrate
on the quasi conservative red detuned far-off-resonance regime.
The system exhibits quasi periodic and chaotic behaviors whose probability can be 
controlled by varying the intensity of the beams.
 The spectral density of the trajectories is used as a chaos signature. An analysis of
permanency times  for chaotic trajectories that visit more than one potential well
reveals a distribution with a long tail.

\end{abstract}
\pacs{42.50.Tx, 37.10.Vz, 06.30.Ka}
\maketitle
\section{Introduction}
The advent of optical cooling and trapping of atoms gave rise to the
experimental and theoretical search of classical and quantum chaotic
 effects in the evolution of atomic center of mass motion
 \cite{levy-cooling,Raizen,tunneling-raizen,argo,levy}.
It was found that the spatial and temporal stochastic
character of spontaneous emission can induce random walks on
the atoms during laser cooling processes  \cite{levy-cooling}. Anomalous
transport properties were discovered, like the fact that both variance and  mean time for
atoms to leave the laser beams could become infinite; the
atoms movement became then a particular class of Levy flights \cite{levy}.

Other interesting effects arise in optical lattices
built from temporally modulated standing waves made using two Gaussian
counter propagating laser beams. If the modulation is chosen to generate an effective
 periodically driven rotor,  the classical dynamics can
 become chaotic and a quantum manifestation is the localization in momentum space
 of the ultra cold atoms \cite{Raizen}.
In  Ref.~\cite{tunneling-raizen}  the quantum dynamical tunneling of ultra cold atoms between
classical islands of stability was also reported; this shows that the presence of chaos
in phase-space can yield substantial consequences on the quantum tunneling rate between
classical regular regions, an effect that has since become
known as chaos-assisted tunneling \cite{tunneling}.

Spatial light modulation to study chaotic dynamics of cold atoms
has also been implemented in the context of
optical billiards. To create an optical billiard, a laser beam
can be deflected at different angles synchronously so that
an arbitrary two-dimensional light pattern is formed in a plane
perpendicular to the optical axis. An additional standing wave
aligned perpendicular to the billiard plane confines the atomic
motion to two dimensions \cite{Milner2001}. This scheme has been
used to investigate the chaotic and regular dynamics of atoms on well-known
billiards \cite{Friedman2001, Andersen2002}. A technique
named echo spectroscopy has been developed to study quantum coherence
in the evolving atomic system \cite{AndersenEcho}. Introducing variable
Gaussian beams waists into this scheme simulates soft-wall billiards
 \cite{Kaplan2001-4}.

In this work we study the semiclassical dynamics of non interacting cold atoms
in optical lattices built from two counter propagating beams with transverse structure.
We numerically show that, even in the simplest case of far-off-resonance quasi
conservative dynamics, the transversal geometry of the beam can be
 used to generate classical chaotic dynamics. This occurs both in the tight binding regime
 (similar to an optical billiard) and in the case that initial conditions
 allow the atomic  transport between different potential wells. All the reported results
  will consider optical lattices with parabolic cylindrical geometry
  although other geometries could also yield similar results.

   In general, the dynamics of cold
 atoms in optical lattices is highly dependent on the optical and
 atomic parameters that nevertheless are feasible to be controlled.
 Dilute atomic samples with predetermined cold atom-cold atom interactions
  can be  used to probe experimentally single-atom  and many-atoms phenomena.
 The detuning between the light frequency and the chosen atomic transition frequency
 is  used to regulate both the relevance of dissipative effects,
 related to spontaneous emission, and the depth and sign of the effective light
 potential affecting the atoms.

It is important to emphasize that structured electromagnetic beams
are experimentally feasible \cite{Julio}, with potential applications
 for the manipulation of cold atomic systems in the semiclassical
 and quantum regimes\cite{allen92}; some of these applications have already been
 implemented  \cite{tabosa,bec-current}.

\section{ Optical lattices with transverse parabolic symmetry}

The electromagnetic (EM) waves with cylindrical symmetry and transverse structure
exhibit many interesting features.  Ideally, these waves have an intensity pattern
invariant under propagation along a given axis that we take as the $z$ axis.
 The best known examples
correspond to Hermite and Bessel waves. The first have Cartesian symmetry and
the latter have circular transverse symmetry \cite{bessel}. Elliptic symmetrical waves
are known as Mathieu waves \cite{mathieu}. The fourth and last separable example,
 has  a transverse structure naturally described in terms
of parabolic cylindrical coordinates $(u,v,z)$ \cite{Lebedev}
\begin{equation}
x + i~y = \frac{1}{2} \left( u + i~v \right)^2,\quad z=z
\end{equation}
where $x$, $y$ and $z$ are the Cartesian coordinates, and
$u\in \left(-\infty, \infty \right)$ and $v \in \left[ 0, \infty \right)$.
 Surfaces of constant $u$ form half confocal parabolic cylinders
that open towards the negative $x$ axis, while the surfaces of
constant $v$ form confocal parabolic cylinders that open in the
opposite direction. The foci of all these parabolic cylinders
are located at $x$=0 and $y$=0 for each $z$ value.
The separability of the wave equation in such a coordinate system,
allows writing the electric and magnetic fields of a monochromatic mode as \cite{para}
\begin{eqnarray}
{\bf E}_{\kappa} &=& - \mathcal{A}^{(TE)}_{\kappa}\partial_{ct}{\mathbb{M}}
 \Psi_{\kappa} - \mathcal{A}^{(TM)}_{\kappa}\partial_{ct}{\mathbb{N}} \Psi_{\kappa}, \\
{\bf B}_{\kappa} &=&\quad \mathcal{A}^{(TE)}_{\kappa}\partial_{ct}{\mathbb{N}}
\Psi_{\kappa} - \mathcal{A}^{(TM)}_{\kappa}\partial_{ct}{\mathbb{M}} \Psi_{\kappa}\nonumber\\
{\mathbb{M}} \Psi_{\kappa} &=& \frac{ \partial}{\partial ct}{\bf\nabla}\times ({\bf e}_z\Psi_\kappa),
-\frac{ \partial}{\partial ct}{\mathbb{N}} \Psi_{\kappa} ={\bf\nabla}\times{\mathbb{M}} \Psi_{\kappa}\nonumber\\
\Psi_\kappa(x,y,z,t) &=&\psi_{\mathfrak{p},k_\bot,a}(x,y) e^{i(k_zz-\omega t)}\nonumber\\
\psi_{\mathfrak{p},k_\perp,a}(x,y) &=&\int_{-\pi}^\pi \mathfrak{A}_\mathfrak{p}(a;\varphi)e^{-ik_\perp(x\cos\varphi + y\sin\varphi)}d\varphi\nonumber\\
\mathfrak{A}_e(a;\varphi) &=& \frac{e^{ia\ln\vert\tan\varphi/2\vert}}{2\sqrt{\pi\sin\varphi}}\nonumber\nonumber\\
\mathfrak{A}_o(a;\varphi) &=& \left\{ \begin{array}{cl} i \mathfrak{A}_e(a;\varphi)&\varphi\epsilon(-\pi,0)\nonumber\\ -i\mathfrak{A}_e(a;\varphi)&\varphi\epsilon(0,\pi). \end{array}\right.\nonumber
\end{eqnarray}
$\mathcal{A}^{(TE)}_{\kappa}$
and $\mathcal{A}^{(TM)}_{\kappa}$ are proportional to the
amplitude of the transverse electric and transverse magnetic modes, and $\Psi_{\kappa}$ is
a solution of the scalar wave equation with $\kappa$ representing the whole set of
numbers that specify a particular mode. That is, the parity $\mathfrak{p}$, the wave vector component along the main axis of propagation $k_z$, the frequency $\omega=c\sqrt{k_z^2 + k_\bot^2}$ and the separation constant $a$ involved in a specific solution of the scalar wave equation.

\begin{figure}[t!]
\includegraphics[height=6cm]{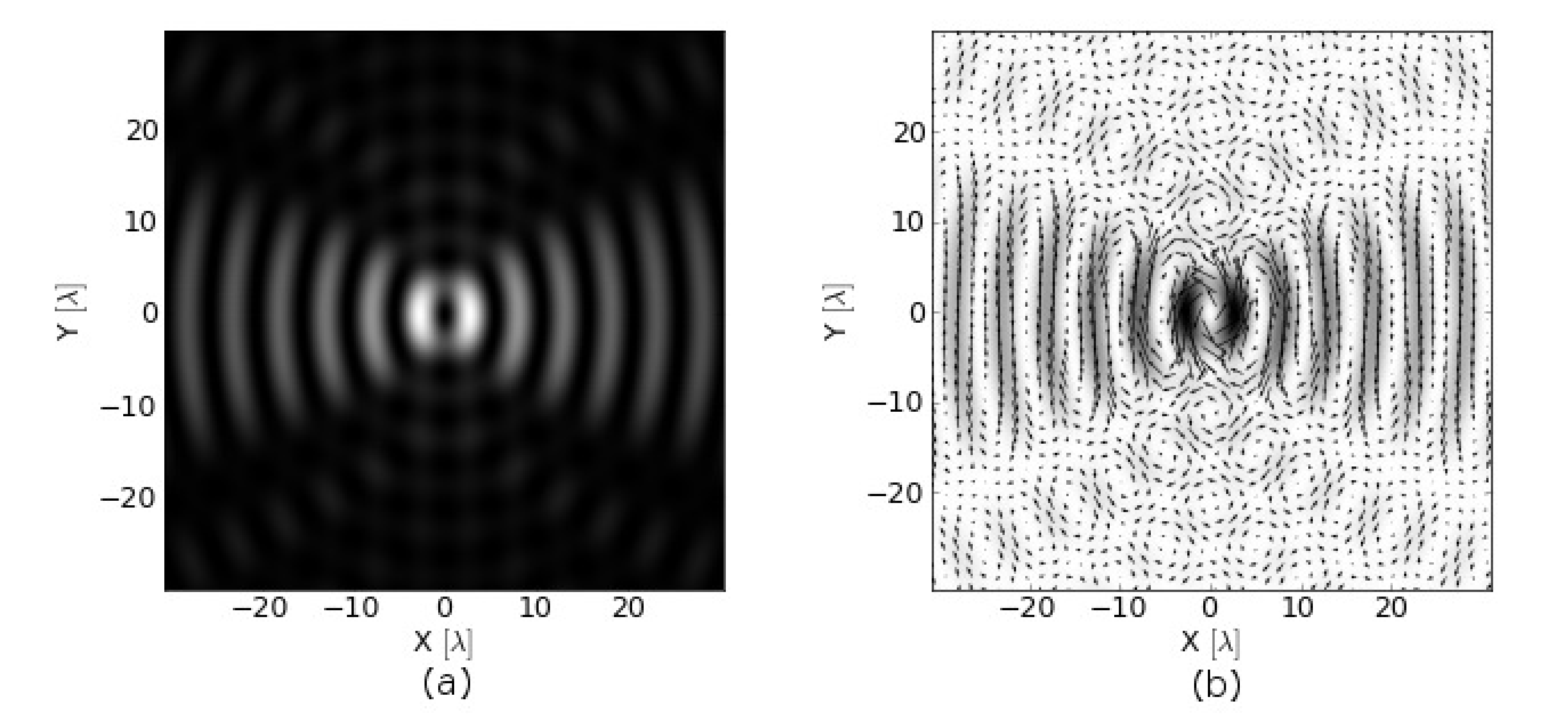}
\caption{(a) Transverse intensity and (b) electric field of a TE electromagnetic wave with parabolic cylindrical symmetry. The wave parameters are $k_z = 0.975 \omega/c$, $a =0$ and the associated
scalar field has even parity. Length is measured in units of the light wavelength $\lambda$. }\label{light}
\end{figure}

 In this work we shall report results for parabolic cylinder electromagnetic
 modes of order $a=0$.  Their intensity and polarization structure are illustrated in Fig.~\ref{light}.
 These modes, for both even and odd parities $\mathfrak{p}$, have been experimentally generated by means of a thin annular slit modulated by the above angular spectra $\mathfrak{A}_\mathfrak{p}(a;\varphi)$ \cite{Lopez2005}.

An optical lattice with parabolic cylinder structure can be generated by the superposition of
two counter propagating beams with that symmetry.
 For simplicity we restrict our study to TE modes with even parity. The structure of the
lattice along the cylindrical symmetry $z$-axis will be that of a sinusoidal standing wave.

\section{Semiclassical force on thermal neutral atoms}
Under standard conditions, the interaction between a two level atom and an EM
wave with a frequency close to resonance has an electric dipole nature with a coupling factor
$$ g^{\pm} = \mu_{12}^{\pm}  ({\bf e}_{x} \pm i {\bf e}_{y}) \cdot {\bf E}.$$
In a semiclassical treatment, the gradient of $g$,
$${\bm\nabla} g= ({\bm\alpha} +i  {\bm \beta}) g,$$defines the force experienced by the atom.
The expression for the average  velocity dependent force, valid for both
propagating and standing beams, is \cite{ashkin}:
\begin{equation}
\langle {\bf f} \rangle = \hbar \tilde\Gamma p^\prime
\Big[[({\bf v}\cdot{\bm\alpha})\frac{1-p}{1+p} +\Gamma/2] {\bm \beta} +
     [({\bf v}\cdot {\bm\beta}) -\delta\omega]            {\bm\alpha}\Big]. \label{eq:force}
\end{equation}
In this expression
\begin{equation}
\tilde\Gamma = \Gamma/[\Gamma(1+p^\prime) +2{\bf v}\cdot{\bm\alpha}[1-p/p^\prime -p][p^\prime/(1+p)]],
 \label{eq:gtilde}
\end{equation}
$\Gamma = 4k^3\vert{\bm\mu}_{12}\vert^2/3\hbar$ is the Einstein coefficient,
 $\delta\omega = \omega -\omega_0$ denotes the detuning
between the wave frequency $\omega$ and the transition frequency
$\omega_0$, $p =2\vert
g\vert^2((\Gamma/2)^2+\delta\omega^2)$ is known as the saturation parameter, linked to the
difference $D$ between the populations of the two levels of the
atom,
 $D=1/(1+p)$, and finally $p^\prime = 2\vert g\vert^2/\vert \gamma^\prime\vert^2$, with
$\gamma^\prime =({\bf v}\cdot{\bm\alpha})(1-p)(1+p)^{-1} +\Gamma/2 +
i[({\bf v}\cdot{\bm\beta})-\delta\omega]$.

The dissipative term $({\bf v}\cdot{\bm\beta})$, associated with a Doppler shift,
as well as other velocity dependent terms in Eq.~(\ref{eq:force}) are expected to be  small
for slow atoms, particularly within the  red detuned far-off-resonance
regime \cite{Miller}. In such a case Eq.(\ref{eq:force}) takes the simpler expression
\begin{equation}
\langle {\bf f} \rangle = \hbar \frac{p}{1+p} \Big[\frac{\Gamma}{2} {\bm \beta} -\delta\omega            {\bm\alpha}\Big]. \label{eq:force2}
\end{equation}
However, we  keep the velocity dependent terms in  our numerical calculations in order to prevent disregarding  potentially relevant effects.

We  consider that the cylindrical symmetry  axis $z$ of the light field
configuration is oriented along the vertical direction and gravity force is included.

\section{Atom semiclassical trajectories}

\begin{figure}[b!]
\includegraphics[height=6cm]{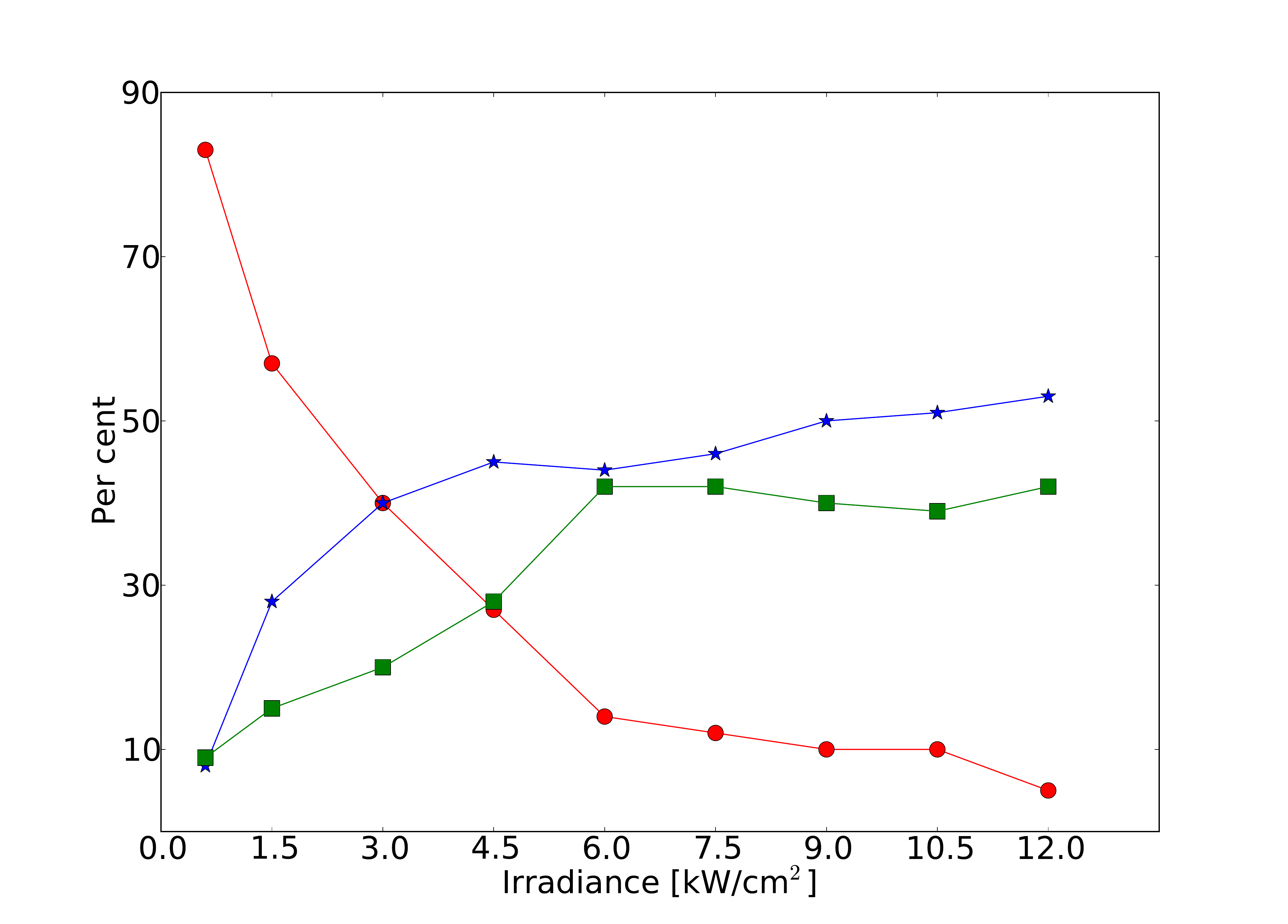}
\caption{Percentage of non trapping (circles), quasiperiodic (stars) and chaotic trapping (squares) trajectories
as a function of the light beam irradiance. The initial conditions for the atomic cloud correspond to random initial positions within a circle of radii $20\lambda$ centered on the axis of the beam and velocities corresponding to a temperature in the range of 2.9-3.1 $\mu$K for the movement in the plane $XY$, perpendicular to the beam, and about
0.2 $\mu$K along the $z$ axis. }
\label{percent}
\end{figure}

\begin{figure}[t]
  \centering
  \subfloat[]{\label{xvycaso88}\includegraphics[width=0.37\textwidth]{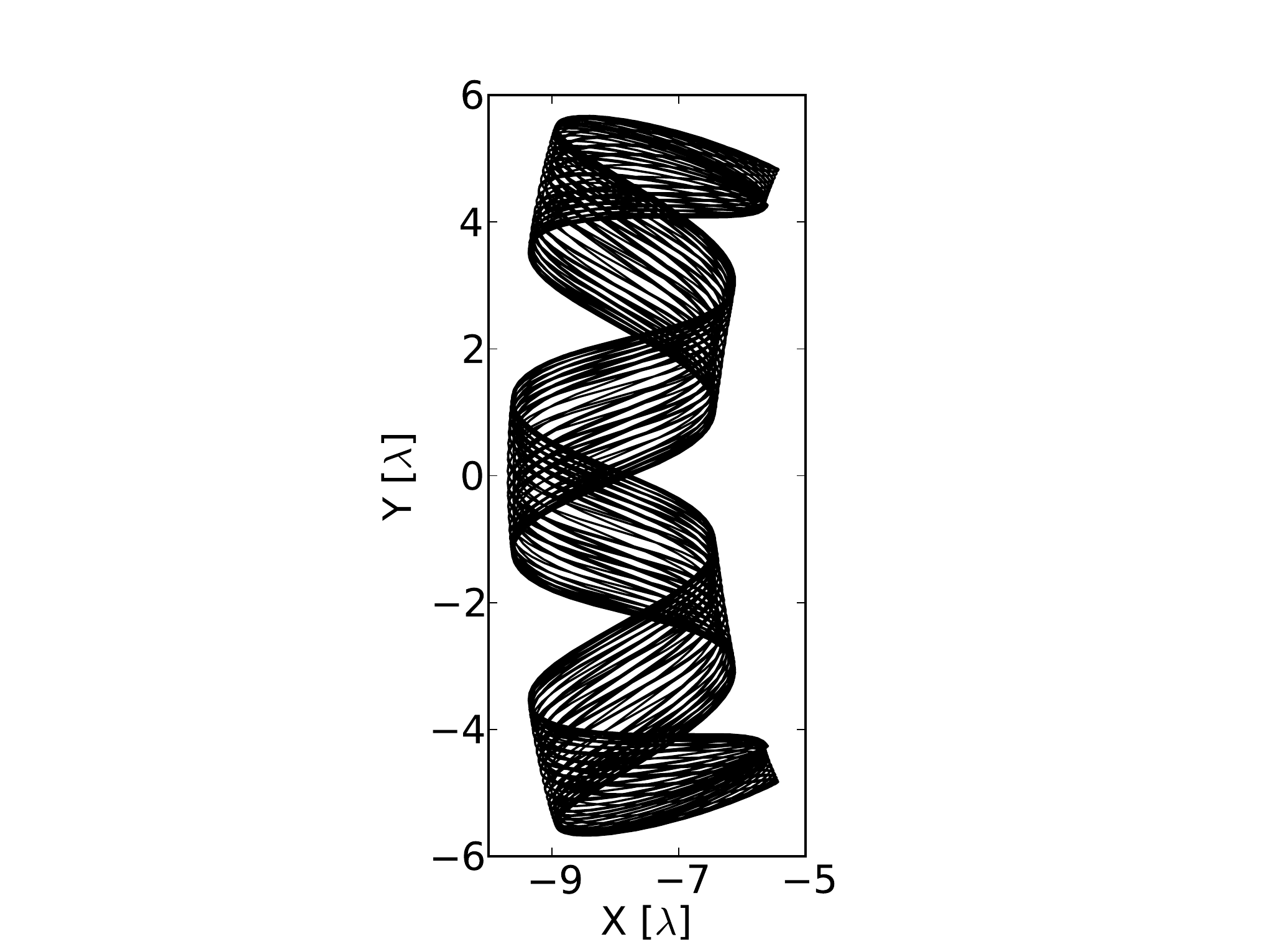}}\hspace{-2.5cm}
  \subfloat[]{\label{xvycaso90}\includegraphics[width=0.37\textwidth]{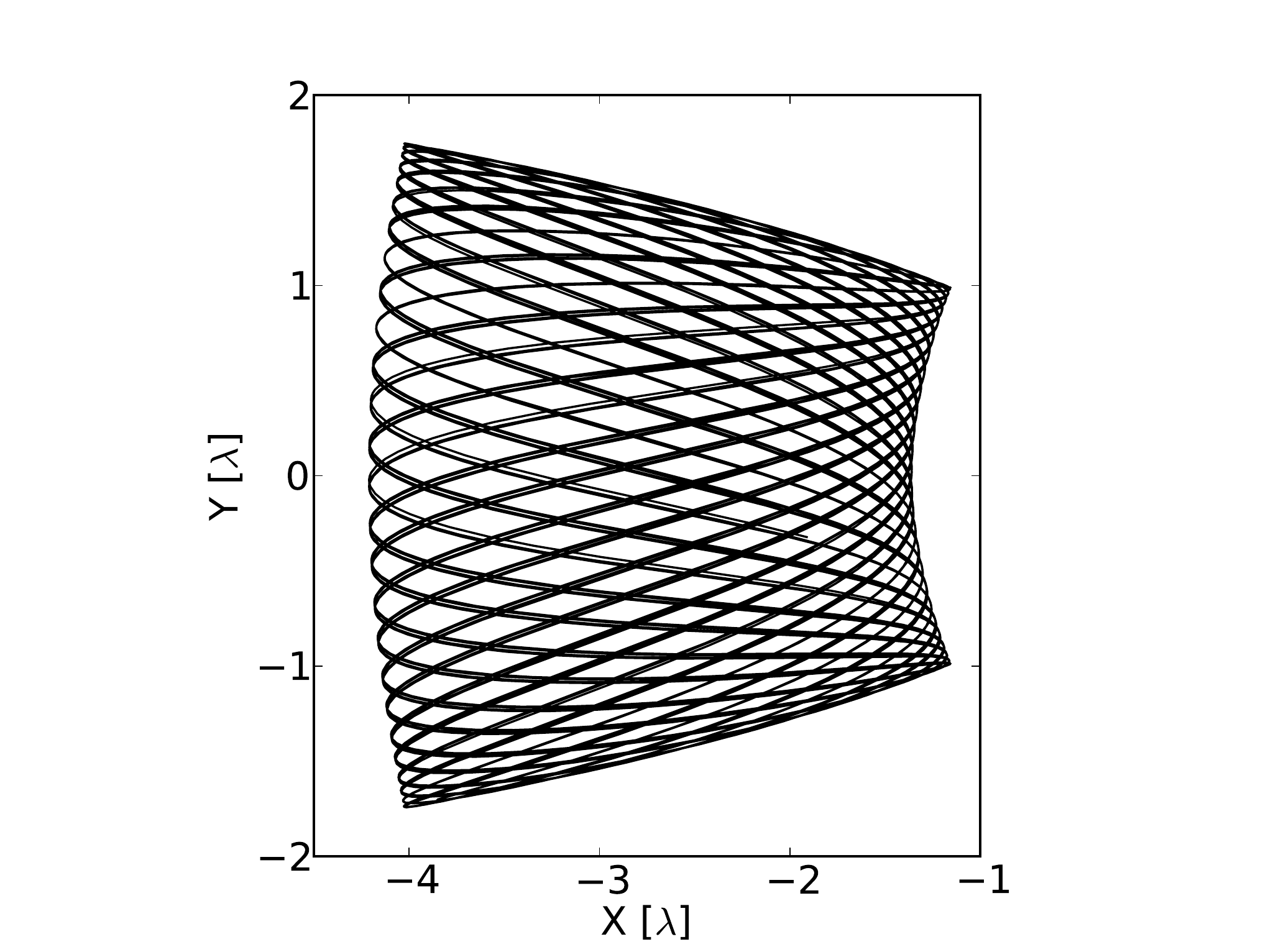}}\hspace{-1cm}
  \subfloat[]{\label{xvycaso51}\includegraphics[width=0.37\textwidth]{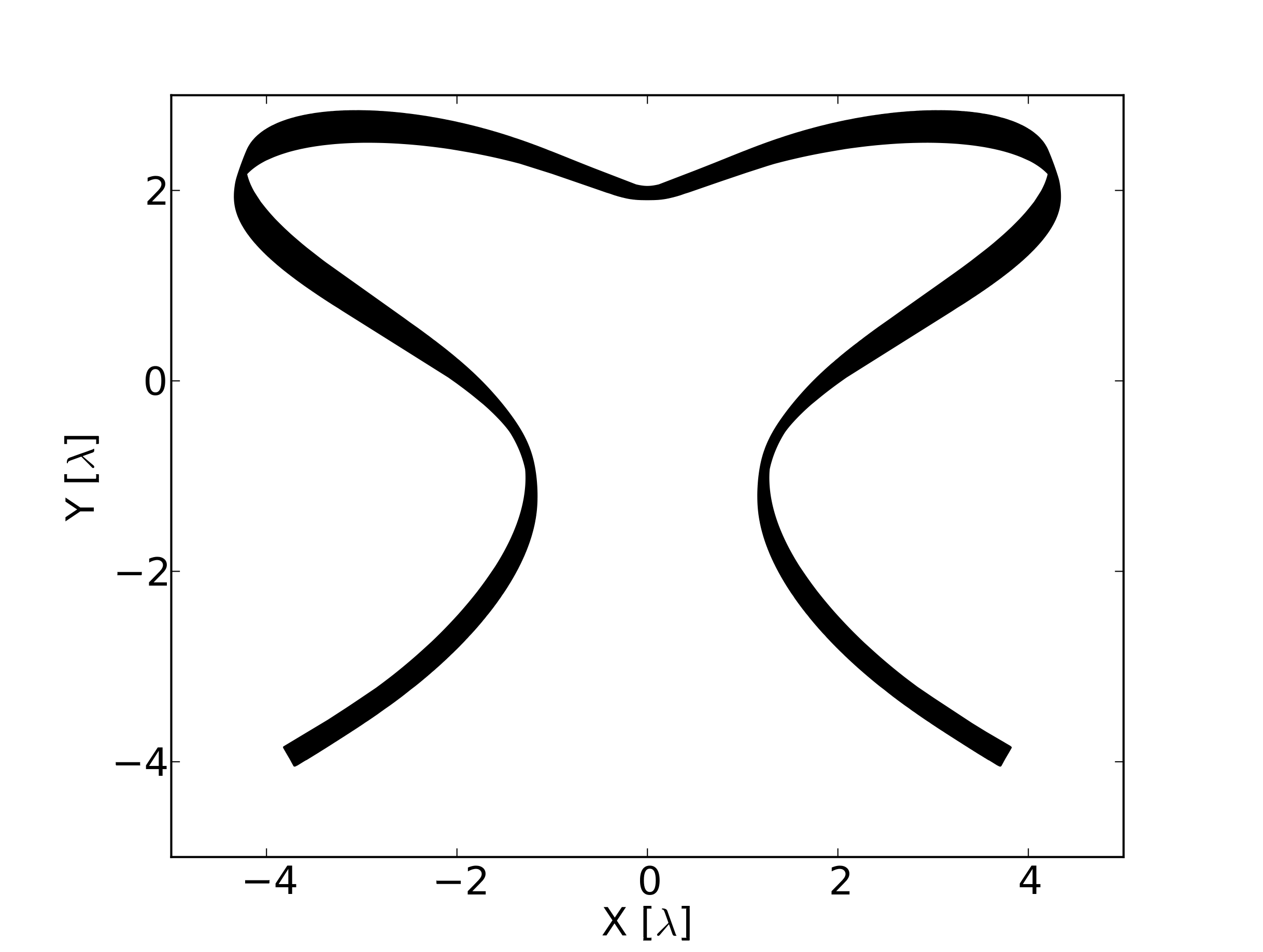}}\\
\caption{ Three illustrative examples of quasiperiodic trajectories of a cold
 atom trapped in a parabolic cylinder lattice: (a) the movement takes place within the
 second lobe to the left of the symmetry axis (see Fig.~\ref{light}); (b) the movement is confined to
 a high intensity lobe closest to the symmetry axis; (c) a typical quasiperiodic trajectory that involves more than one optical lobe. For clarity the time interval plotted is a few percentage ($\sim 10\%$) of the 1.25 $\times$ $10^8$ $\Gamma^{-1}$  time considered in the numerical calculation. Length is measured in units of the light wavelength $\lambda$. }
\label{cuasi123}
\end{figure}

\begin{figure}[b]
\includegraphics[height=6cm]{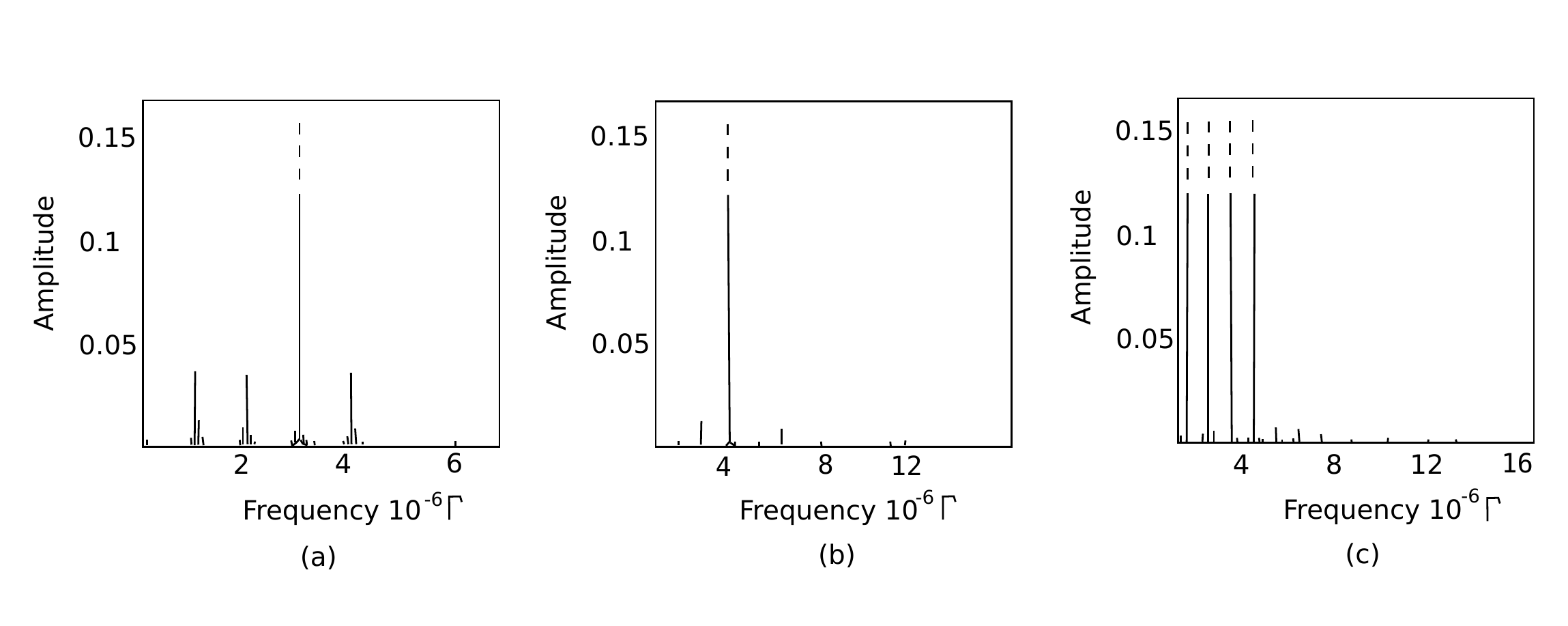}
\caption{Power spectra of the  coordinate $x(t)$ for the three
quasiperiodic trajectories shown in Fig.~\ref{cuasi123} showing the typical well defined isolated
periodicities expected for a regular dynamics. Only the low frequency end of the spectra
is shown. The frequency is measured cycles per unit time $\Gamma^{-1}$.}
\label{pwcuasi123}
\end{figure}
The numerical simulations \cite{de} consider a TE laser beam detuned 67 nm to the red of the $5~^2S_{1/2}$ -
$5~^2P_{1/2}$ transition at $795$~nm of $^{85}\text{Rb}$ with irradiance
in the range $\sim 0.5-22$ kW/ cm$^2$. As natural length and time units, we take
the laser wavelength and the inverse of the Einstein coefficient $\Gamma$,
which is $3.7 \times 10^{7}$s$^{-1}$ for the state $5~^2P_{1/2}$ of $^{85}\text{Rb}$.

Given an irradiance, initial conditions for a cloud of a hundred non interacting atoms are generated randomly within a circle of radii $20\lambda$ centered on the axis of the beam. The random velocities correspond to a temperature in the range of 2.9-3.1 $\mu$K for the movement in the plane $XY$, perpendicular to the beam, and about
0.2 $\mu$K along the $z$ axis. The anisotropy in the initial velocities  enhances the effects of
the nontrivial transverse structure since it "freezes" the z-direction degree of freedom.
A mechanism for achieving initial anisotropic velocities at low temperatures could be based 
in the previous use of a dissipative optical lattice with an anisotropic modulation of
the laser-atom interaction parameters \cite{sanchez-placencia}.

In Figure \ref{percent} we illustrate the proportion of non trapped, regular trapped and chaotic trapped
trajectories as a function of the irradiance. For instance
for an irradiance of $\sim$4.5 kW/cm$^{2}$ about $27\%$ of the atoms are
not trapped by the lattice and escape with their route either almost confined in a
transverse plane  to the beam or with their velocity almost parallel to it.
The remaining atoms exhibit a trapped chaotic ($28\%$) or quasi periodic motion ($45\%$).
An atom will be considered trapped if for times lower
than $1.25\times 10^8 \Gamma^{-1}$ it remains at a distance to the axis lower
than $80\lambda$ while, in the axial direction, it remains within at most $5\lambda$
of its initial $z$ value. An irradiance of $\sim$ 2 kW/cm$^{2}$ is required for
trapping $\sim 50\%$ of the atoms. Notice however, that for an irradiance as low as
$\sim$0.5 kW/cm$^{2}$ there is a high probability of observing
chaotic trajectories within the trapped ones.
For an irradiance of $\sim 6$ kW/cm$^2$ $\sim 82\%$ of the atoms are trapped
by the lattice, about $\sim 35\%$ have a chaotic motion and $\sim 47\%$ a regular one.
In the following,  the cases that illustrate the atoms dynamics refer to the latter irradiance.

In most of the regular trajectories, the atoms are confined to a single lobe of the lattice (Fig.~\ref{cuasi123}a,~\ref{cuasi123}b) although less than 10$\%$ of the observed regular trajectories can involve two lobes (Fig.~\ref{cuasi123}c) or more than one well along the $z$-direction.
We could not find regular trajectories visiting more than two lobes in the XY plane.
The quasiperiodic character of the motion can be easily verified by evaluating the
power spectra  of any of the components of the position or
velocity vectors as a function of time as shown in Fig.~\ref{pwcuasi123}.

\begin{figure}[t]
 \centering
  \subfloat[]{\label{xvycaso36}\includegraphics[width=0.37\textwidth]{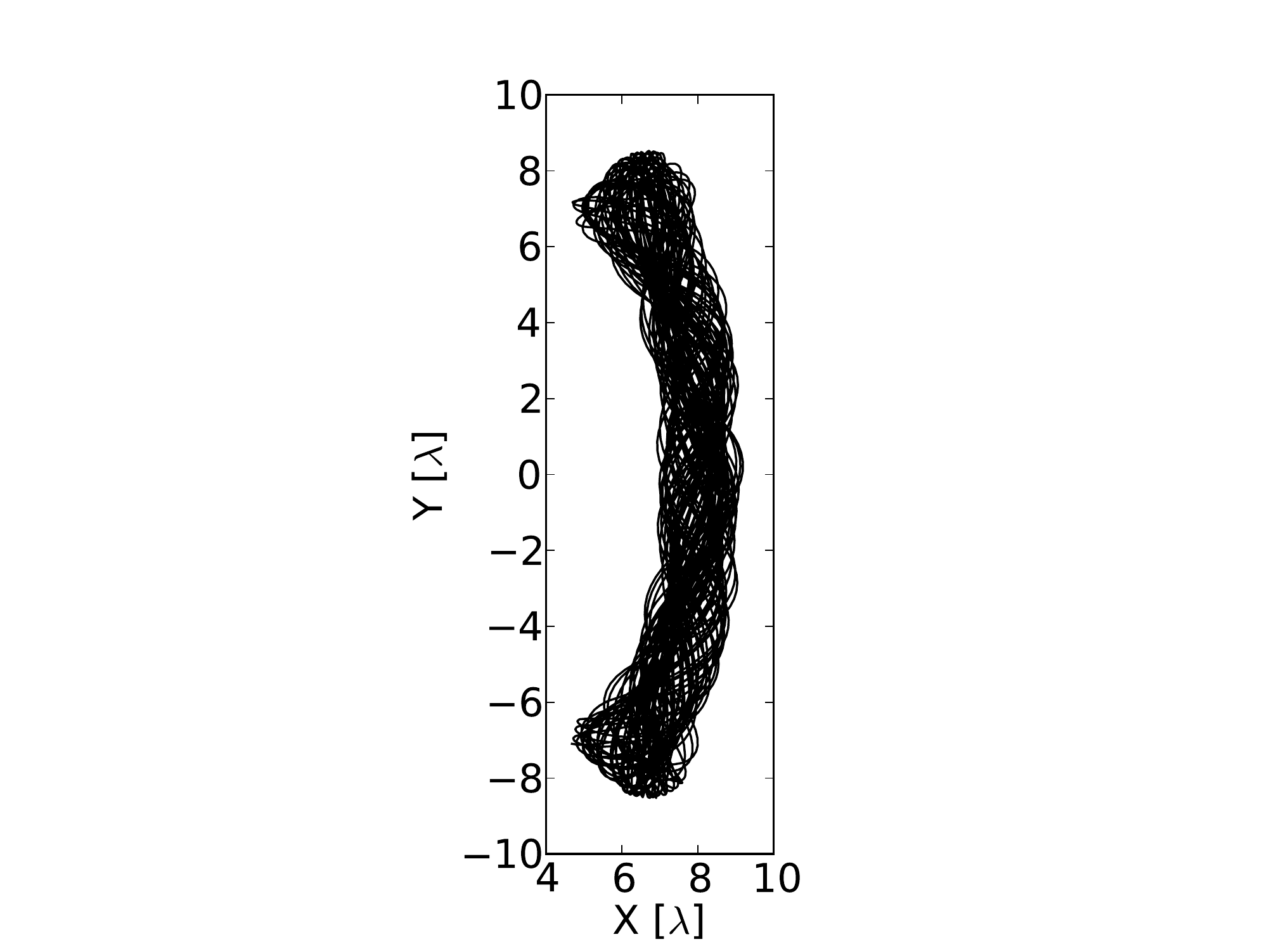}}\hspace{-1cm}
  \subfloat[]{\label{xvycaso87}\includegraphics[width=0.3\textwidth]{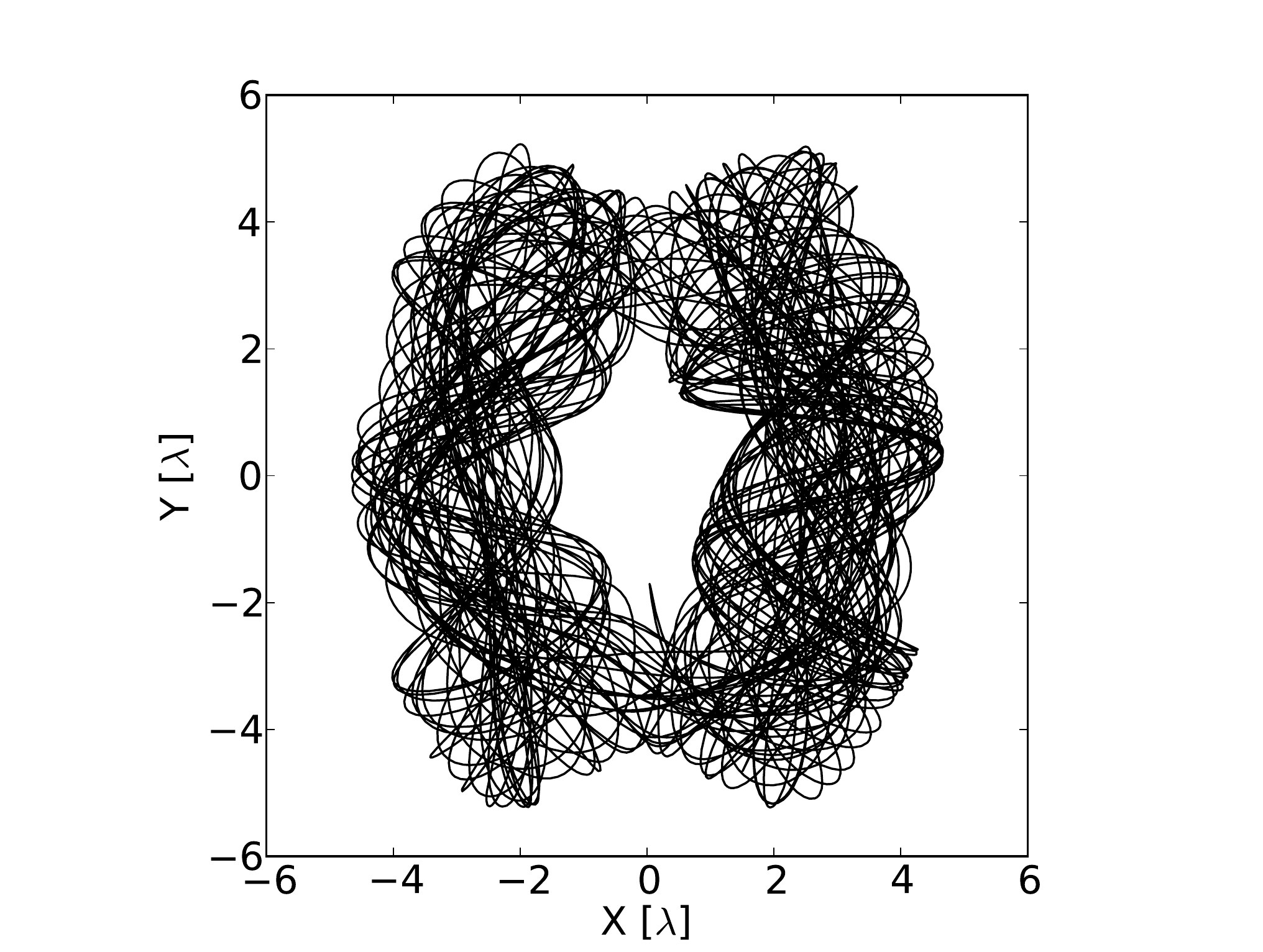}}
  \subfloat[]{\label{xvycaso99}\includegraphics[width=0.37\textwidth]{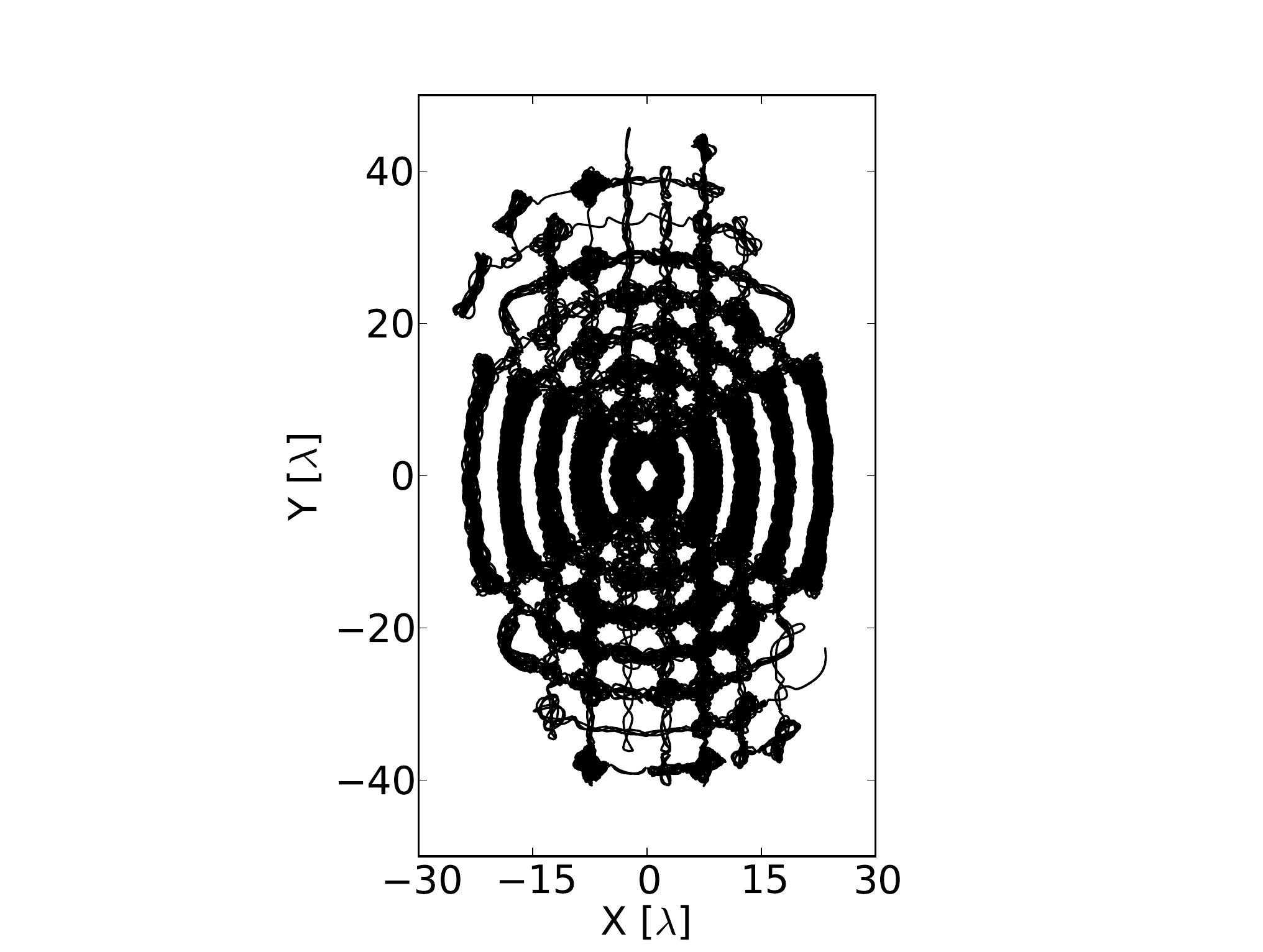}}\\
\caption{Three illustrative examples of chaotic trajectories where:
 (a) the movement takes place within the
 second lobe to the right of the symmetry axis (see Fig.~\ref{light}); (b) the movement is confined to
 the two highest intensity lobes closest to the symmetry axis; (c)  the trajectory practically covers the accessible configuration space in the optical lattice.
For clarity the time interval plotted is a few percentage ($\sim 10\%$) of the 1.25 $\times$ $10^8$ $\Gamma^{-1}$  time considered in the numerical calculation. Length is measured in units of the light wavelength $\lambda$. }
\label{chaotic123}
\end{figure}

\begin{figure}[b]
\includegraphics[height=6cm]{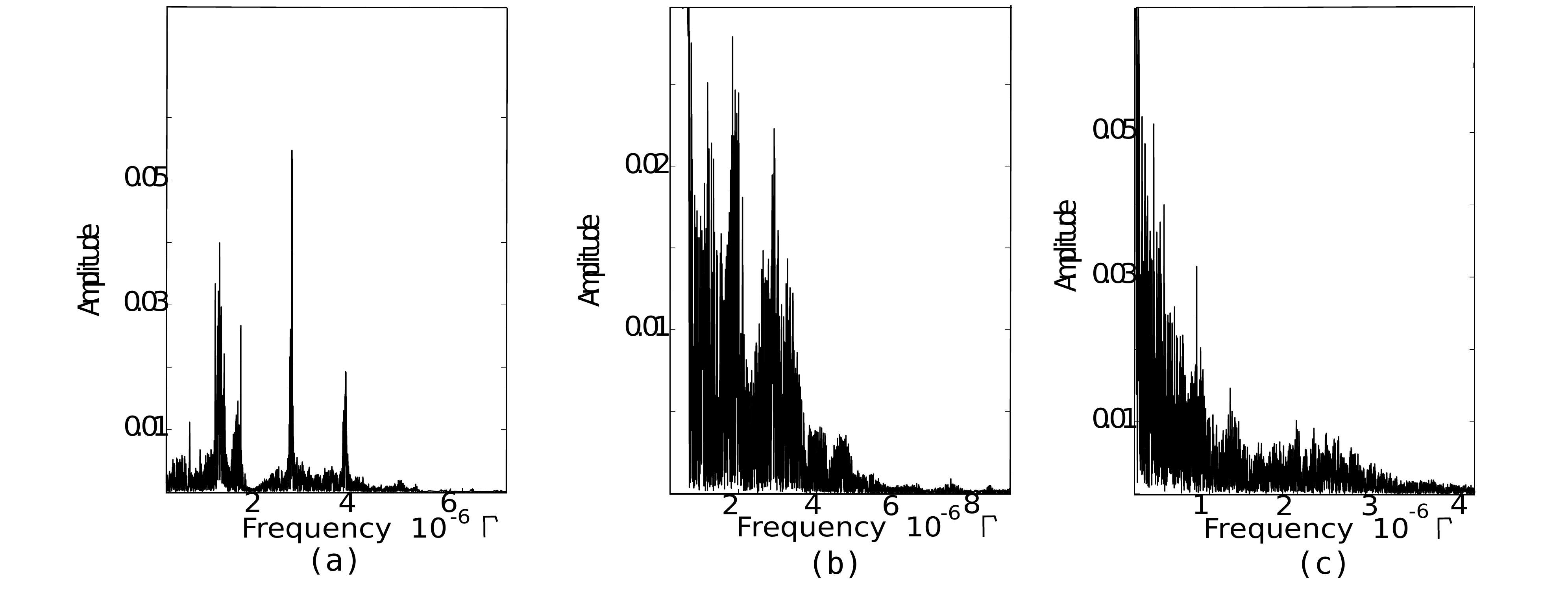}
\caption{Power spectra  of the coordinate $x(t)$ for the
three chaotic trajectories shown in Figure \ref{chaotic123}. The broad band spectra
is a signature of chaotic motion. Only the low frequency end of the spectra
is shown. The frequency is measured in cycles per unit time $\Gamma^{-1}$.}
\label{pwchaotic123}
\end{figure}

As for chaotic trajectories, for the initial conditions we considered, they give rise to
confined motion within one wavelength in the $z$ direction and transversal
motion in either one, two or several lobes in the labyrinths created by
the optical lattices in the $XY$ plane, as illustrated in Fig.~\ref{chaotic123}.
In the most common chaotic trajectories, the atom remains in a lobe for certain time and
then goes to another lobe of the lattice in a very irregular way that
practically covers the accessible configuration space in the optical lattice, Fig.~\ref{chaotic123}c.

For atom trajectories involving many lobes, the lobes are visited in an irregular order 
and an atom can visit the same lobe several times. For Fig.~\ref{chaotic123}c, the atom visits most of the lobes in
any interval $(t_0,t_0+\Delta t)$ of duration $\Delta t=1.25\times10^7\Gamma^{-1}$
irrespective of the initial time $t_0$.

In Fig.~\ref{pwchaotic123} the power spectra of the $x$ coordinate of the trajectories shown in
Fig.~\ref{chaotic123} is presented. The broad band structure of those spectra is a signature of
the trajectories chaotic character.

\begin{figure}[b]
\includegraphics[height=6cm]{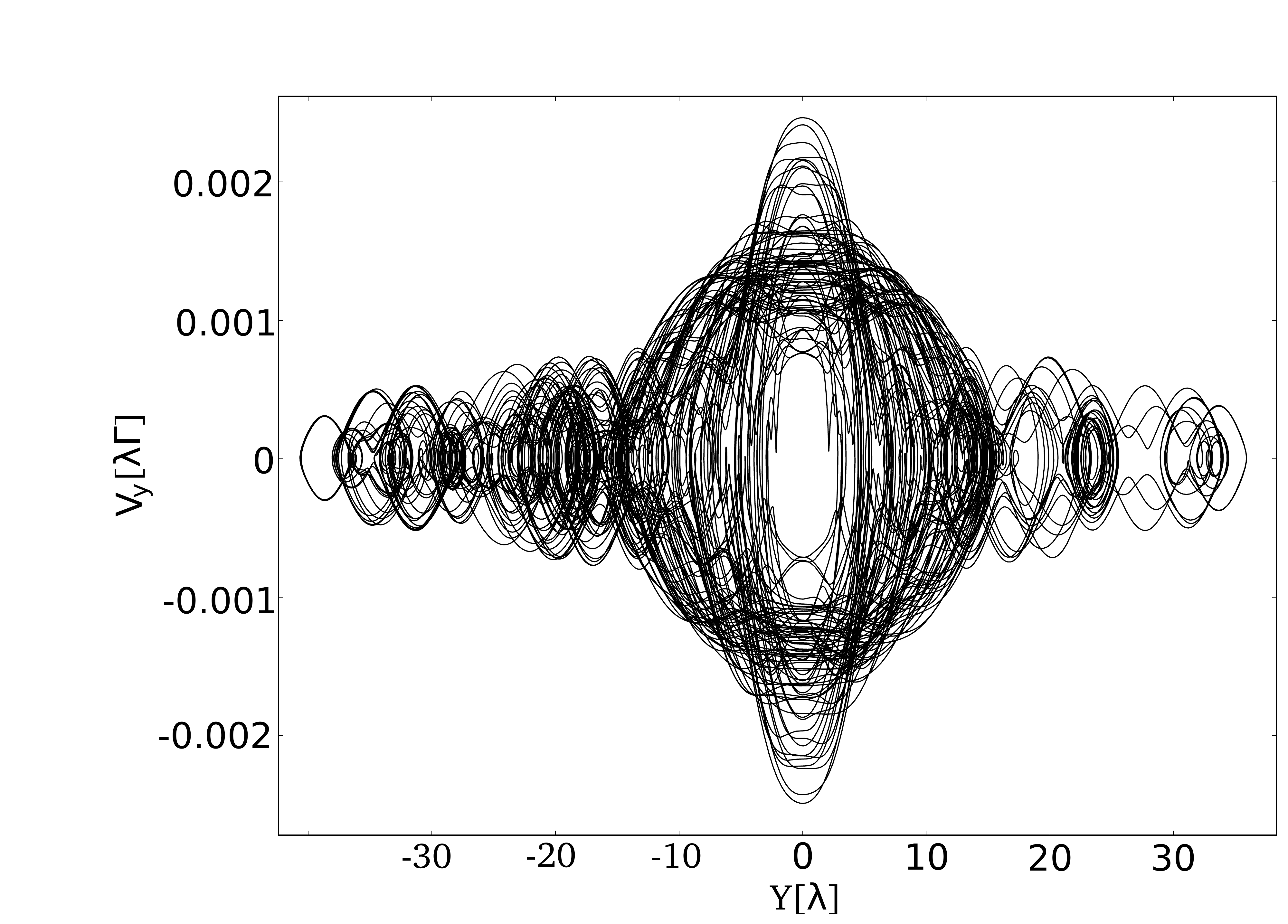}
\caption{Trajectory in the velocity $V_y$ versus position $y$ phase space
 plane for the path 4c. }\label{yvy}
\end{figure} 
   
 Looking for other signatures of chaos,
the phase space trajectory can also be also analyzed. In Fig.~\ref{yvy} we illustrate
such trajectories in ($y$,$V_y$) space. Both Fig.~\ref{chaotic123}c and Fig.~\ref{yvy} point to a
dense covering of the accessible phase-space established by the
light labyrinth and the atom initial conditions.
\begin{figure}[b]
\includegraphics[height=6cm]{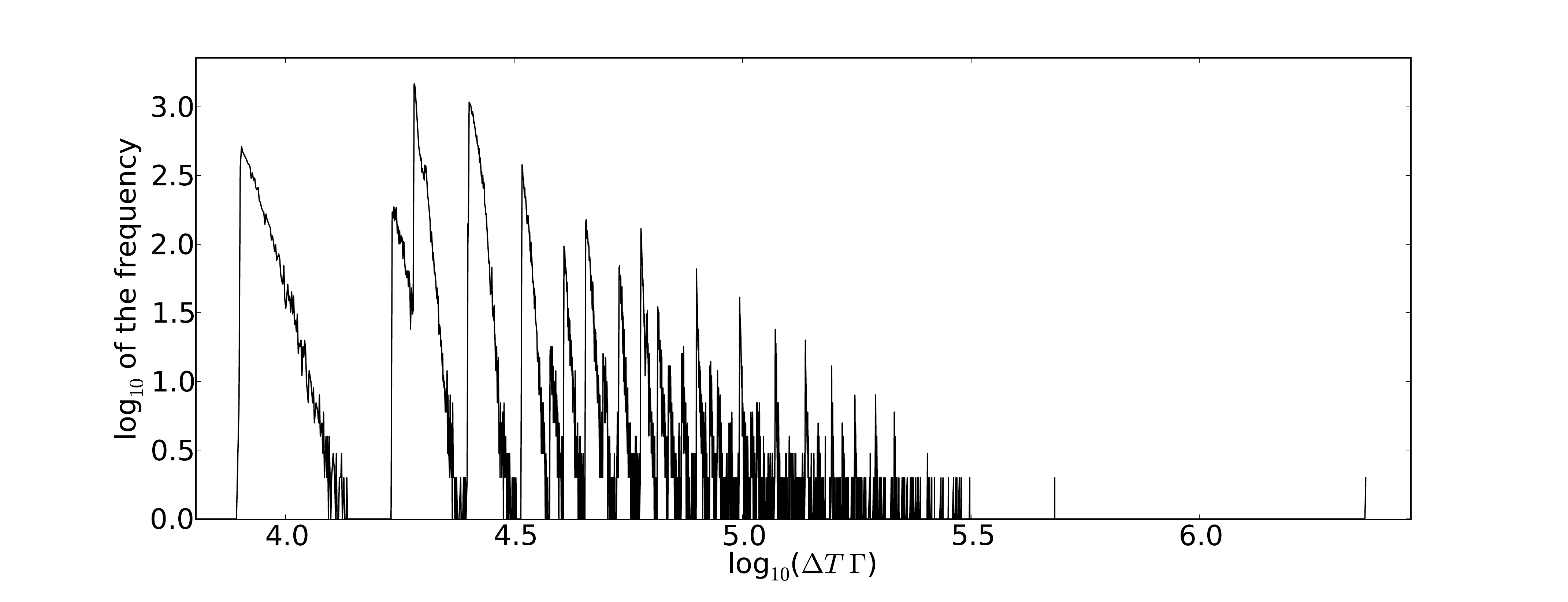}
\caption{Log--log plot illustrating the statistics of the time an atom expends in an optical lobe before
jumping to another optical lobe for a chaotic trajectory. The sampling was taken for the trajectory 
illustrated in Fig \ref{chaotic123}b after an analysis that considered a total time $T_F = 2.5\times 10^8 \Gamma^{-1}$  
in which 82951 transitions took place. The frequency corresponds to the number of events on which the 
atom expended a time $[T,T+ 50\Gamma^{-1}]$ within a lobe.
The long tail is a signature of a statistics where the momenta is not well defined. In this figure, the logarithm of zero has been mapped to zero.}\label{levy}
\end{figure}

In Fig.~\ref{levy}, we illustrate the distribution of the permanency time $T_{perm}$ within a lobe  
in a log-log plot for the trajectory showed in Fig.~\ref{chaotic123}b.
The sampling was taken   considering a total  evolution time $T_F = 2.5\times 10^8 \Gamma^{-1}$  
in which the atom transits 82951 times from one to the other lobe. 
Analysis with shorter total evolution time (we studied $T_F$ in the interval $[0.5\times 10^8 \Gamma^{-1},2.5\times 10^8 \Gamma^{-1}]$) gave frequency distributions of $T_{perm}$ with the same structure; that is
(i)the same characteristic minimum time an atom expends within a lobe (a time naturally determined by the initial kinetic energy); (ii) local very stepped maxima followed by strong decays which are then followed by the next maxima;
(iii) the longest permanency time $T_{longest}$ has a very small frequency;  
 (iv) if the longest permanency time is not considered the distribution of the local maxima 
  is approximately linear, which would then lead to a power law statistics for that maxima;
  (v) $T_{longest}$ increases as $T_F$ increases. 
   This indicates to us that the momenta of the distribution is not well defined, like in a
  Levy distribution.

\begin{figure}[b]
\centering
\subfloat[]{\label{transit_160}\includegraphics[width=0.4\textwidth]{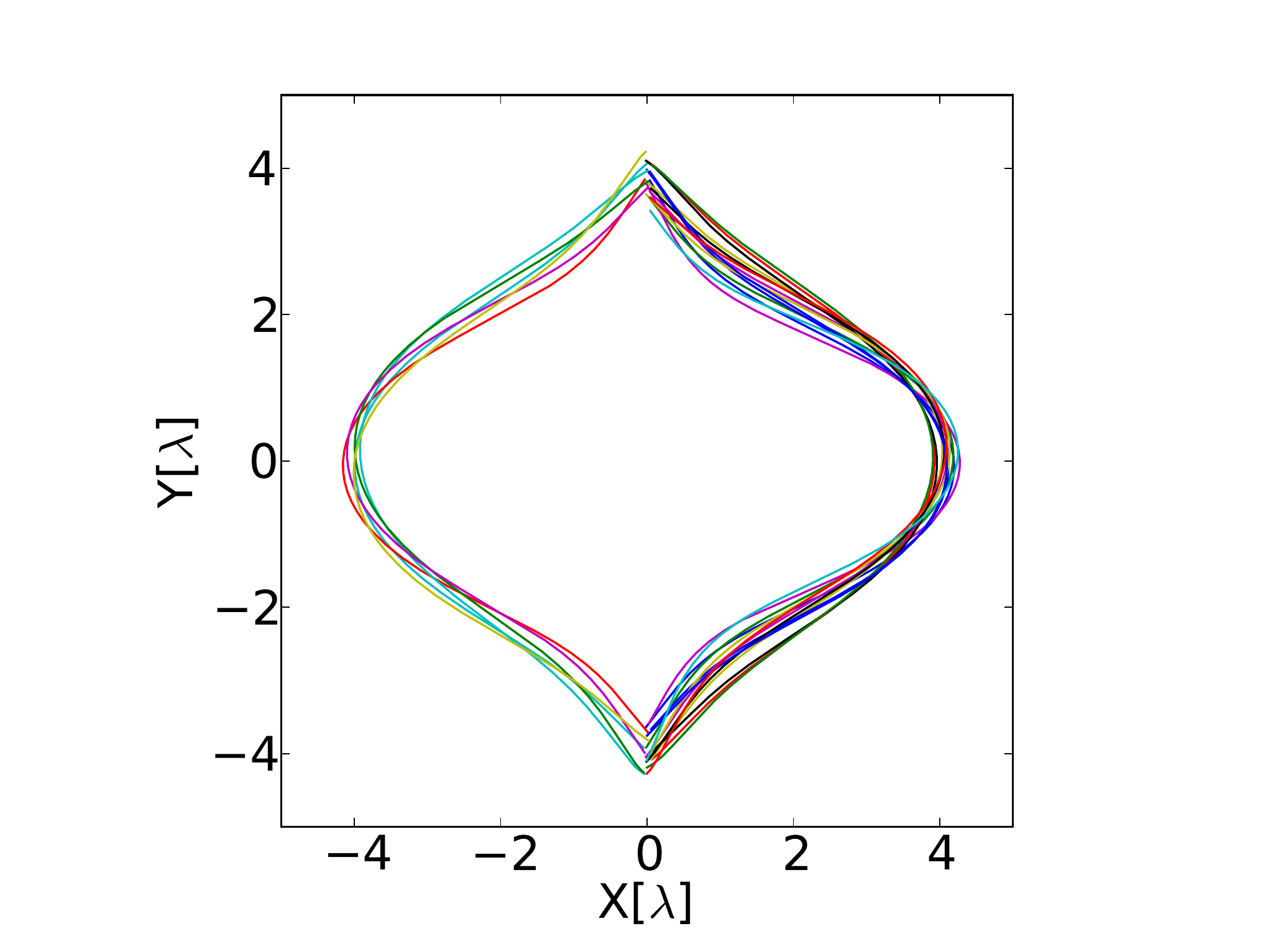}}
\subfloat[]{\label{transit_650}\includegraphics[width=0.4\textwidth]{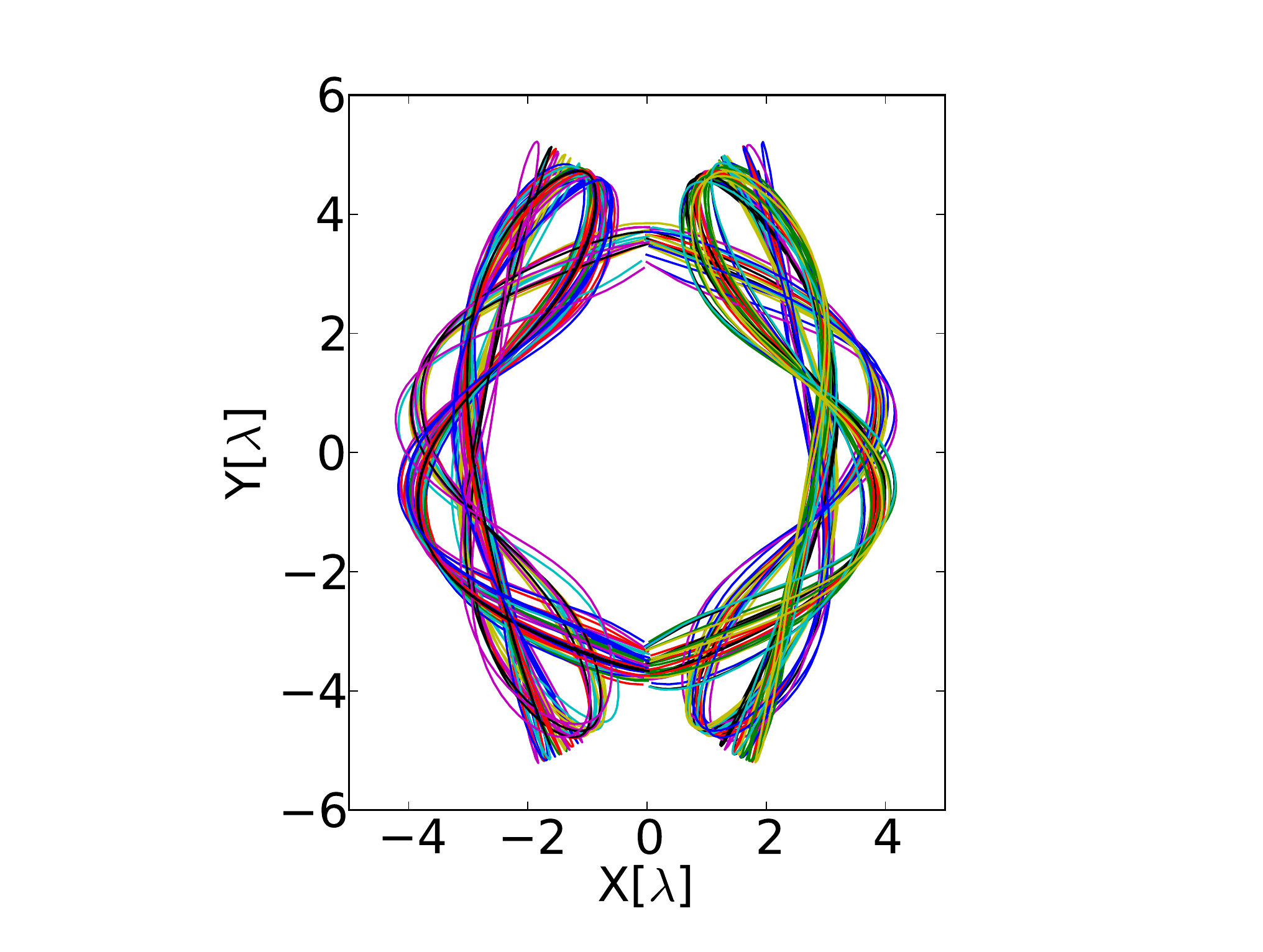}}
\caption{(Color on line) Illustrative partial trajectories during the time an atom remains within a lobe
for the chaotic trajectory shown in Fig.~\ref{chaotic123}b. They were classified according to the time permanency $T_{perm}$
so that in (a) $T_{perm}=8000\pm 1000\Gamma^{-1}$,  (b) $T_{perm}=32500\pm 2000\Gamma^{-1}$}\label{part1}
\end{figure}

\begin{figure}[b]
\centering
\subfloat[]{\label{transit_384}\includegraphics[width=0.4\textwidth]{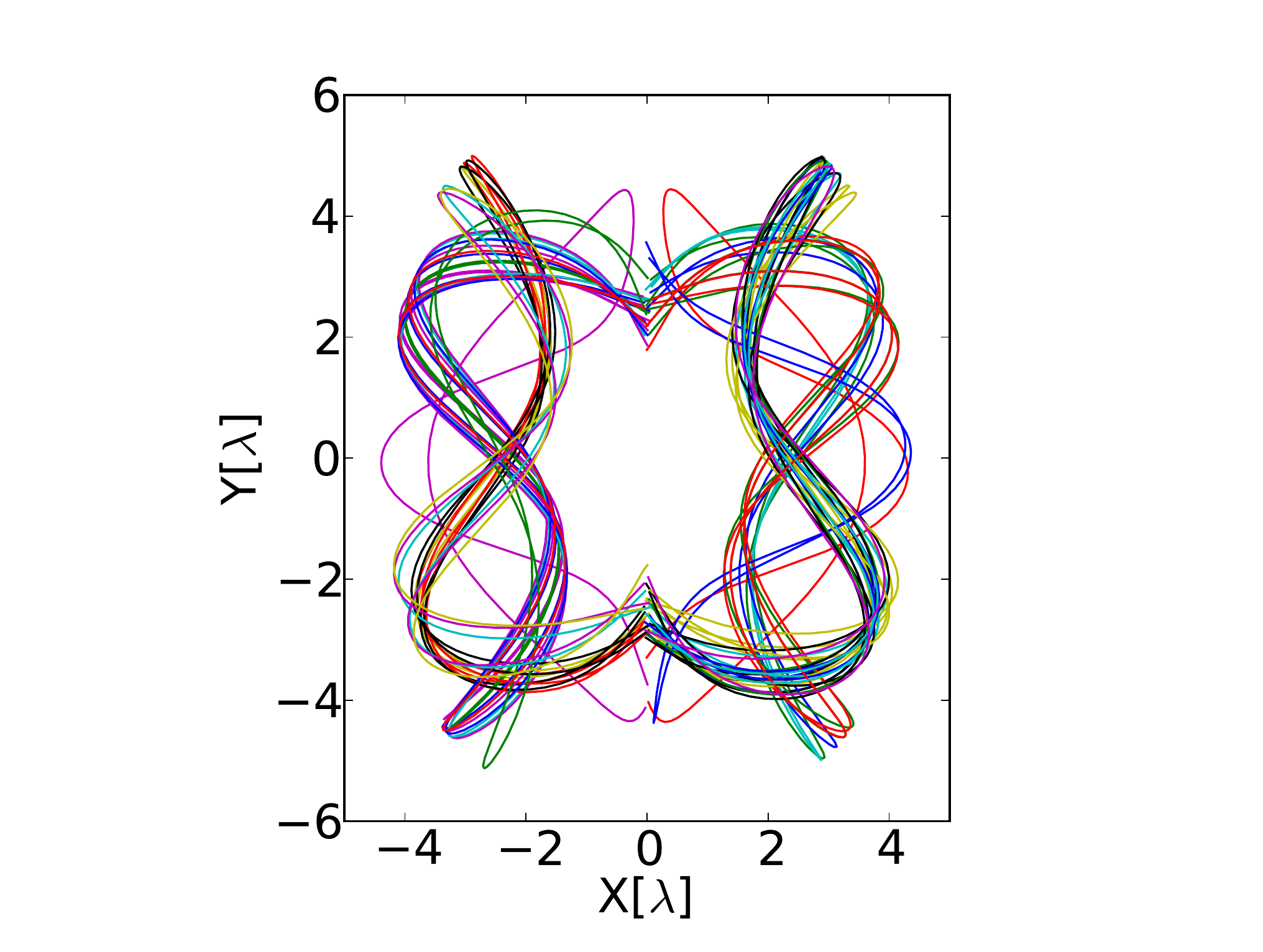}}
\subfloat[]{\label{transit_400}\includegraphics[width=0.4\textwidth]{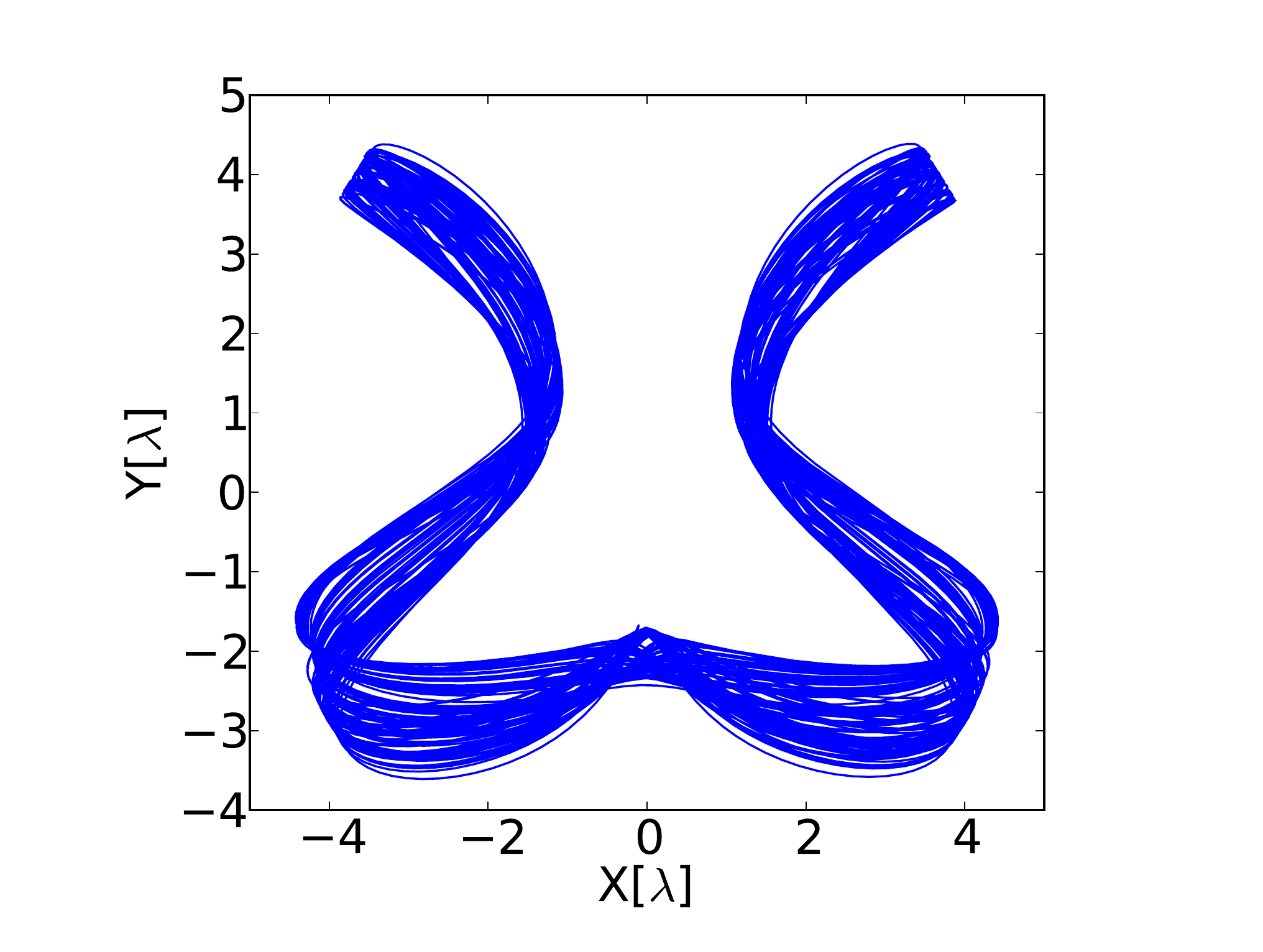}}
\caption{(Color on line) Illustrative partial trajectories during the time an atom remains within a lobe
for the chaotic trajectory shown in Fig.~\ref{chaotic123}b with a permanency time $T_{perm}=19200\pm 1500\Gamma^{-1}$.
There is a non negligible probability that the atom passes through this type of partial trajectories
in consecutive times as shown in (b).}\label{part2}
\end{figure}

   Partial trajectories with nearby values of the permanency time $T_{perm}$
 exhibit similarities as illustrated in Figs.~\ref{part1} and \ref{part2}.  Fig.~\ref{part1}a  refers to the
 trajectories with $T_{perm}$ equal to the  minimum time an atom must expend in a lobe. For them,
 the atom enters through one of the potential maxima and leaves through the other one. 
 In most cases the partial trajectories associated with the local maxima of the distribution, 
 correspond to the ones using the minimum time necessary to transit within a lobe touching a certain
 number of times the boundary to the other lobe, as illustrated in Fig.~\ref{part1}b.
  In Fig.~\ref{part2}, we illustrate the type of trajectories with major frequency in the distribution of
  $T_{perm}$. In this interesting case, we observe the presence of two kind of partial trajectories that
  go across the lobe entering through one boundary and going out through the same boundary;
 just one of them almost arrives to the second boundary. The other class of trajectories has a
  non negligible probability of being followed by an almost identical trajectory under reflection
  when the atom crosses to the other lobe. This is clearly illustrated in Fig.~\ref{part2}b where consecutive
  partial trajectories of this kind lead to a partial trajectory very similar to the quasiperiodic
case shown in Fig.~\ref{cuasi123}c.

\section{Conclusions}
We have numerically shown that optical lattices with transverse structure
allow the design of potentials yielding non trivial dynamics. For chaotic trajectories,
like the one shown in Figure \ref{chaotic123}c,  the atom has a non zero probability
to be found at any point within its accessible  region of  phase space. This behavior
is similar to that expected for the quantum description of the atom motion. In this
case the atom wave function will necessarily lead to a position
probability density that mimics the optical structure, being non zero in any of the lobes.
We expect that  for ultra cold atoms chaos-assisted bidimensional tunneling will take place.

Under the proposed set-up, one can  notice that the fraction of trapped atoms with chaotic trajectories diminishes as the irradiance is lowered, Fig.~\ref{percent}. It is also important to emphasize
that, for intermediate values of the irradiance,  most of the trapped chaotic trajectories involve more than one lobe and that the distribution of time permanencies  within a lobe  is highly structured and with a  long tail. 
Finally, for high irradiances, the proportion of chaotic trajectories saturates  and there is a high presence of quasiperiodic trajectories.  In this last regime, the atom is highly confined in the $z$-direction and 
tends to move inside a single optical potential lobe. Since the detuning is large, dissipative effects are expected to be small, thus, in the high intensity regime, the system dynamics is very similar to that of  two-dimensional billiards with a topology determined by the parabolic symmetry of the light beam. It is  known\cite{quasi-bill} that this kind of dynamical systems have a large set of quasiperiodic solutions. We consider this the key to understand the saturation effect illustrated in Fig.~\ref{percent}.

Notice that the use of a propagation invariant EM wave is not mandatory for
the qualitative results we found. The use of other beams like the Gaussian
parabolic are expected to yield similar results for the transverse motion of atoms in the
focusing plane.

We also assumed anisotropic initial velocities. If the mean initial z-velocity 
were comparable with the transverse velocities, one would expect that the
atoms motion would take place in multiple quasi bidimensional configuration spaces since the atom
would visit several of the sinusoidal potential wells in the z-direction. However, 
the projection of the trajectories in a given transversal plane would be expected 
to be qualitatively similar to those described above.

For the geometries we are studying, blue detuned beams, due to the numerous routes of escape provided by 
the dark light zones, in general do not lead to trapped trajectories. Nevertheless this scattering process
deserves a detailed study on its own, both classically and quantum mechanically.

\end{document}